\documentclass{aa}  
\usepackage{graphicx}
\usepackage{txfonts}
\usepackage{hyperref}
\usepackage{adjustbox}
 \usepackage{lineno}
 \usepackage{textgreek}
\usepackage{xcolor}
\usepackage{siunitx}

\usepackage{graphicx}
\usepackage{txfonts}
\begin{document}

\title{The cosmological analysis of X-ray cluster surveys: VI.  Inference based on analytically simulated observable diagrams}
\titlerunning{Simulation-based inference of cosmology with galaxy clusters and machine learning}

   \author{M. Kosiba\inst{1, 2, 3}
          \and
          N. Cerardi\inst{4}
          \and
          M. Pierre\inst{2}
          \and 
          F. Lanusse\inst{2}
          \and \\
          C. Garrel\inst{5}
          \and
          N. Werner\inst{1}
          \and
          M. Shalak\inst{6}
          }

               

    \institute{Department of Theoretical Physics and Astrophysics, Faculty of Science, Masaryk                    University, Kotl\'a\v rsk\'a 2, Brno, 611 37, Czech \\
               Republic
        \and
            Universit\'e Paris-Saclay, Universit\'e Paris Cit\'e, CEA, CNRS, AIM, 91191, Gif-sur-Yvette, France
        \and
            Dipartimento di Fisica, Universit\`{a} degli Studi di Torino, via Pietro Giuria 1, I-10125 Torino, Italy
        \and
            Universit\'e Paris Cit\'e, Universit\'e Paris-Saclay, CEA, CNRS, AIM, F-91191, Gif-sur-Yvette, France
        \and
             Max Planck Institute for Extraterrestrial Physics, Giessenbachstrasse 1, D-85748 Garching, Germany
        \and
            Laboratoire Univers et Th\'ories, Observatoire de Paris, CNRS,
            Universit\'e PSL, F-92190 Meudon, France
        }

   \date{Received; accepted}


  \abstract
   {The number density of galaxy clusters across mass and redshift has been established as a powerful cosmological probe, yielding important information on the matter components of the Universe. Cosmological analyses with galaxy clusters traditionally employ scaling relations, which are empirical relationships between cluster masses and their observable properties. However, many challenges arise from this approach as the scaling relations are highly scattered, maybe ill-calibrated, depend on the cosmology, and contain many nuisance parameters with low physical significance.}
   {
   In this paper, we use a simulation-based inference method utilizing artificial neural networks to optimally extract cosmological information from a shallow X-ray survey, solely using count rates, hardness ratios, and redshifts. This procedure enables us to conduct likelihood-free inference of cosmological parameters $\Omega_{\mathrm{m}}$ and $\sigma_8$.
   
   }
   {To achieve this, we analytically generate several datasets of 70\,000 cluster samples with totally random combinations of cosmological and scaling relation parameters. Each sample in our simulation is represented by its galaxy cluster distribution in a count rate (CR) and hardness ratio (HR) space in multiple redshift bins. We train
   Convolutional Neural Networks (CNNs) to retrieve the cosmological parameters from these distributions. We then use 
   neural density estimation (NDE) neural networks to predict the posterior probability distribution of $\Omega_{\mathrm{m}}$ and $\sigma_8$ given an input galaxy cluster sample.
   }
   {Using the survey area as a proxy for the number of clusters detected for fixed cosmological and astrophysical parameters, and hence of the Poissonian noise, we analyze various survey sizes. The 1 $\sigma$ errors of our density estimator on one of the target testing simulations are 1\,000\,deg$^2$: 15.2\% for $\Omega_{\mathrm{m}}$ and 10.0\% for $\sigma_8$; 10\,000\,deg$^2$: 9.6\% for $\Omega_{\mathrm{m}}$ and 5.6\% for $\sigma_8$. We also compare our results with a traditional Fisher analysis and explore the effect of an additional constraint on the redshift distribution of the simulated samples.
   }
   {We demonstrate, as a proof of concept, that it is possible to calculate cosmological predictions of $\Omega_{\mathrm{m}}$ and $\sigma_8$ from a galaxy cluster population without explicitly computing cluster masses and even, the scaling relation coefficients, thus avoiding potential biases resulting from such a procedure.
   }


   \maketitle
   


\section{Introduction}

Galaxy clusters are the largest virialized systems in the Universe, sitting on the peaks of the primordial density fluctuation field \citep[e.g.][]{Bahcall1988}. The mass distribution and number density of galaxy clusters across space and redshift are governed by the cosmological parameters of the Universe, making them potentially powerful cosmological probes \cite[e.g.][]{Eke1996}. Clusters in X-rays are less prone to projection effects than galaxy over-densities in the optical \citep[e.g.][]{Bhatiani2022} and provide key physical information on the intra-cluster medium (ICM), which is a good tracer of the underlying dark matter distribution \citep[e.g.][]{Holland2015,Fischer2023}.  

However, cluster cosmology is usually considered a delicate enterprise, given that the cluster mass, a key parameter that enters the cosmological analysis, is not a directly observable parameter. Rather than using mass proxies, like the X-ray temperature, luminosity, or gas mass, our approach is to forward model a set of observed properties. More specifically, we consider the distribution of the count rates (CR) and hardness ratios (HR) observed by the XMM-Newton satellite in several redshift bins (X-ray Observable Diagrams, XOD). 
The main idea is that the CR and HR values are analogous quantities to the X-ray fluxes and the temperatures of the galaxy clusters. Together with the redshift information, these quantities thus carry both physical and cosmological information.
This method, ASpiX, has proven an efficient way of performing cluster cosmological analyses \citep{Clerc2012}. It enables easy modelling of the cluster selection function and has the merit of cleaning several potential systematic uncertainties that are otherwise cosmology-dependent. Moreover, it enables the inclusion of all detected clusters in the analysis because HR measurements require much fewer photons than measuring a temperature.

To date, the method has been tested on simulations \citep{Pierre2017, Valotti2018} and used to extract the cosmological constraints from the XMM-XXL cluster sample (\cite{Garrel2022}, Garrel et al. (2024), submitted). All these applications used a Markov Chain Monte Carlo (MCMC) algorithm to infer the cosmological parameters and related uncertainties. Moreover, cosmology-dependent priors derived from the same cluster sample were imposed on the cluster scaling relations, linking the mass with the cluster luminosity, temperature, size, and associated scatter values.

The present paper is the sixth of the series investigating the potential of the ASpiX concept. Our goal here is to use a deep learning approach to explore to what extent we can infer cosmology without any prior knowledge of the cluster evolutionary physics, i.e., on the coefficients of the scaling relations. 
We start by creating XODs for a wide range of $\Omega_{\mathrm{m}}$ and $\sigma_8$ values, assuming random values for the scaling relation coefficients, which will be handled as nuisance parameters: this allows us to produce an unbiased simulated XOD sample.  
We then train a sequential neural posterior estimation (SNPE) \citep{Papamakarios2016} network, a novel machine learning (ML)-based technique of the simulation-based inference family (SBI), to infer the cosmological parameters $\Omega_{\mathrm{m}}$ and $\sigma_8$ from the XODs.
We consider several survey sizes independently to estimate the effect of the shot noise on cluster counts, thus on the representation of the XODs and subsequently on the inferred cosmological parameters.

This paper is structured as follows: Section\,\ref{sec:Data} describes the formalism used to create the XOD sample. Section \,\ref{sec:Method} describes the SNPE method. We present our main results in section \,\ref{sec:Results}. In section \,\ref{sec:Discussion}, we discuss the impact of various factors on our cosmological predictions: the size of the training set for the neural networks, the effect of imposing a prior on the cluster redshift distribution, and of decreasing the number of free parameters by assuming self-similar evolution.  We discuss here also what steps would need to be taken when using this method on a real observed XOD of our Universe. We conclude in section \,\ref{sec:Conclusions}.

In this paper, we shall consider uniform XMM-Newton extragalactic surveys of various sizes, all performed with an exposure time of 10 ks at any sky position. We take WMAP9 \citep{Hinshaw2013} as our fiducial cosmology.



\section{Creating the set of X-ray Observable Diagrams}
\label{sec:Data}
Given an XMM-Newton extragalactic survey area, size, and exposure time, the XOD is a summary statistic of the entire detected cluster population. Namely, a 3D representation of the HR, CR, and $z$ observed parameters.  CR is the total (MOS1+MOS2+pn) count rate in the [0.5-2] keV band, HR is the ratio of the number of counts in the [1.0-2.0]/[0.5-1.0] bands; while measurement errors can be attributed to CR and HR, we consider that spectroscopic redshifts are available with negligible uncertainty. The XODs thus represent the cluster number counts in the CR-HR-$z$ parameter space.

To create the XOD set for the cosmological inference, one needs to populate the cluster multi-parameter space (cosmology and cluster physics) in a way that is tractable for this 3-dimensional summary statistics. In the XOD formalism, cluster properties have to be expressed in terms of CR, HR, and z. To compute the first two quantities, one must know, as a function of cluster mass and redshift, the cluster luminosity and temperature. The fact that clusters follow scaling relations is an intrinsic property, physically motivated by the theory of structure formation and observationally verified. Consequently, we make use of scaling relations to create our XOD set. For this, we use simple power laws because, in the survey regime that we are modeling (10ks XMM observations), we collect an average of 200 photons per cluster \citep{Lieu2016}. This does not allow measurements of T (and L) with sufficient accuracy to implement broken power laws. Similarly, we cannot quantify the effect of the cool core, as we mostly detect the cluster central region. Both effects can be considered as integrated in the very wide range assumed for the scaling relation coefficients.


\subsection{Cosmological model and galaxy cluster mass function}
To simulate the X-ray cluster population, we assume a flat $\Lambda$CDM cosmology ($w=-1$) and the \cite{Tinker2008} halo mass function (HMF). This, multiplied by the comoving volume element, provides us with the number of galaxy clusters (dn) formed per unit mass (dM) per redshift bin (dz) per unit solid angle (d$\Omega$). The ranges of considered $\Omega_{\mathrm{m}}$ and $\sigma_8$ values are given in Table \ref{table:table_parameters}.



\subsection{Scaling relation formalism}
The next step is to attribute to each halo an X-ray luminosity and a temperature; this is done through scaling relations. We adopt the formalism described by \cite{Pacaud2018}:
\begin{equation}
    \frac{M_{500}}{M_0\times10^{13}M_{\odot}h^{-1}} = \Bigg(\frac{T_{300 \mathrm{kpc}}}{1\mathrm{keV}}\Bigg)^{\alpha_{MT}} E(z)^{\gamma_{MT}} 
\end{equation}

\begin{equation}
    \frac{L_{500}}{L_0\times10^{41}\mathrm{erg} \mathrm{s}^{-1}} = \Bigg(\frac{T_{300 \mathrm{kpc}}}{1\mathrm{keV}}\Bigg)^{\alpha_{LT}} E(z)^{\gamma_{LT}} 
\end{equation}

\noindent where $L_{500}$ is the rest-frame luminosity in the [0.5-2] keV band, within $R_{500}$;
$T_{300 \mathrm{kpc}}$ stands for a generic measure of the temperature within a radius of 300 kpc. 
The ranges adopted for the six free parameters (normalization, slope, evolution) are given in Tab.\,\ref{table:table_parameters}.
Following \cite{Pacaud2018}, we encapsulate all the scatter in the L-T relation, i.e. $\sigma_{LT}$ = 0.67, and $\sigma_{MT}$ = 0. The scatter in the scaling relations is fixed to a value obtained from standard survey observations \citep{Pacaud2018}. We could have also let this parameter free, but we feel that this would unnecessarily increase the degeneracy with the other scaling relation parameters, on which we do not put any constraint.

\begin{table}[h]
\caption{Sampling range for parameters used to simulate the CR-HR-$z$ diagrams. The first two are the cosmological parameters of interest for the present study;  the latter six define the scaling relations. For all sets of simulated XOD, the eight parameters are drawn at random, assuming a uniform probability within the given ranges. The adopted ranges are $\pm 50\%$ of the fiducial values; except for $\gamma_{LT}$ for which the current uncertainties are much larger.}           
\label{table:table_parameters}      
\begin{tabular}{cccc}
    \hline \hline
    coefficients       & central value    & min value    & max value  \\
    \hline
    \hline
    $\Omega_{\mathrm{m}}$    & 0.279 	            & 0.1395              & 0.4185                  \\
    $\sigma_{8}$    & 0.821 	            & 0.4195              & 1.2315                  \\
    $M_{0}$         & 2.6 	                & 1.3                 & 3.9                     \\
    $\alpha_{MT}$   & 1.67 	                & 0.835               & 2.505                   \\
    $\gamma_{MT}$   & -1.0 	                & -1.5                & -0.5                    \\
    $L_{0}$         & 8.24                  & 4.12                & 12.36                   \\
    $\alpha_{LT}$   & 3.17 	                & 1.585               & 4.755                   \\
    $\gamma_{LT}$   & 0.47 	                & 0                   & 1.2                     \\

\hline
\end{tabular}
\end{table}


\subsection{X-ray observable properties}
After each $(M, z)$ halo is assigned a luminosity and a temperature, we derive the corresponding observed total XMM-Newton count rates in the three [0.5-2], [1-2], and [0.5-1] keV bands of interest; this is done using the APEC plasma model folded with the XMM-Newton responses for the three EPIC detectors as described e.g. in \cite{Garrel2022}.


\subsection{XMM-Newton survey design and selection function}
In the present study, we consider several (hypothetic) survey sizes ranging from 1\,000 deg$^2$ to infinity (see Sec.\,\ref{sec:Noise_and_survey_area} and Sec.\,\ref{Survey area - Noise}). This allows us to study the impact of the shot noise in the XODs: the larger the survey, the larger the number of clusters in the dCR/dHR/dz bins, hence the smaller the statistical fluctuations in each bin. 

To simulate the survey selection function, we assume a simple CR cut. This allows us to significantly simplify the modeling of the selection process, which in reality depends on both the CR and the cluster apparent size (\cite{Pacaud2006}).  This hypothesis has no incidence for the purpose of the present study. We set the cut to CR = 0.02 c/s, which corresponds to $\sim 200$ counts for a 10 ks XMM-Newton exposure; this is a safe limit inspired by the XXL survey statistics \citep{Adami2018} and yields for the fiducial cosmology and the \cite{Pacaud2018} scaling relations around 4 clusters / deg$^2$.


\subsection{Construction of the XOD sample and the d$n$/d$z$ selection}
\label{sec:XOD_construction_and_dndz}
\begin{figure*}[!th]
\begin{center}
    \includegraphics[height=7cm,width=15cm,angle=0]{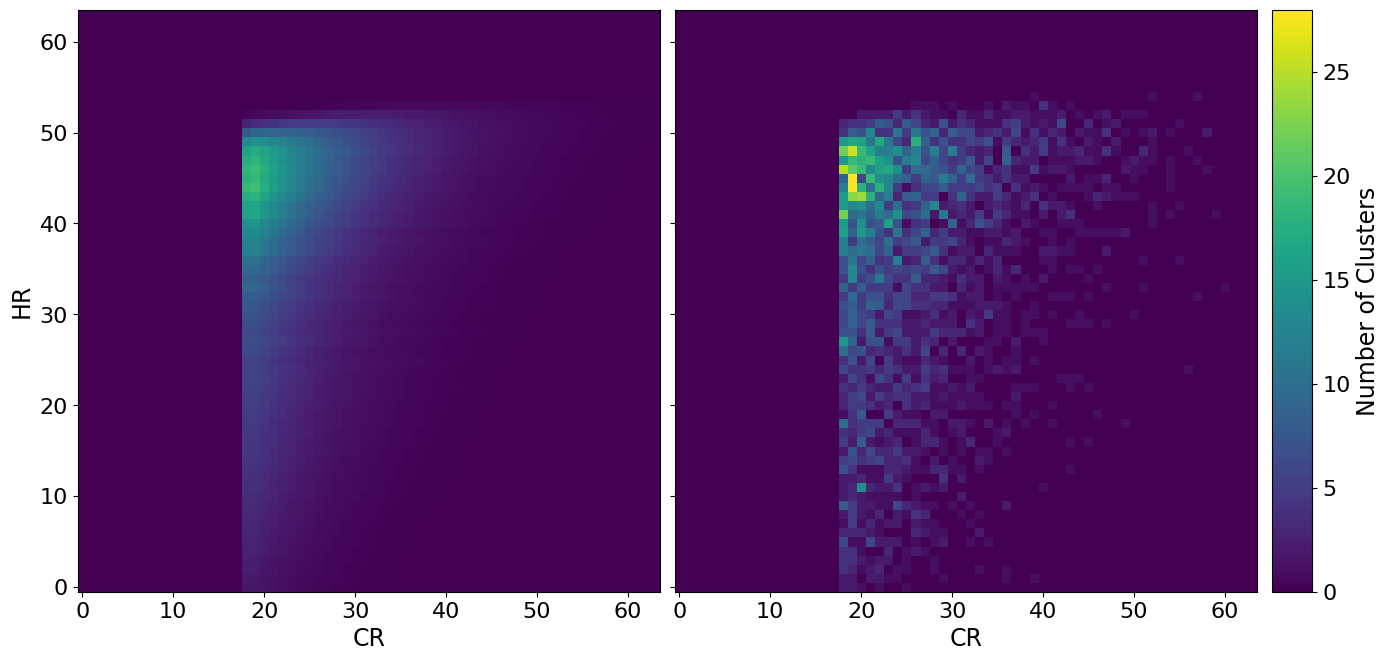}
    \caption{Left: Fiducial XOD for a 1\,000 deg$^2$ survey area (simulated for central values of Table \ref{table:table_parameters}), integrated over the redshift dimension. The XOD has a 64$\times$64$\times$10 resolution. Right: a particular Poisson realization of the left-hand-side image.}
    \label{fig:XOD_examples_64x64}
\end{center}
\end{figure*}

\begin{figure*}[!th]
\begin{center}
    \includegraphics[height=7.0cm,width=15cm,angle=0]{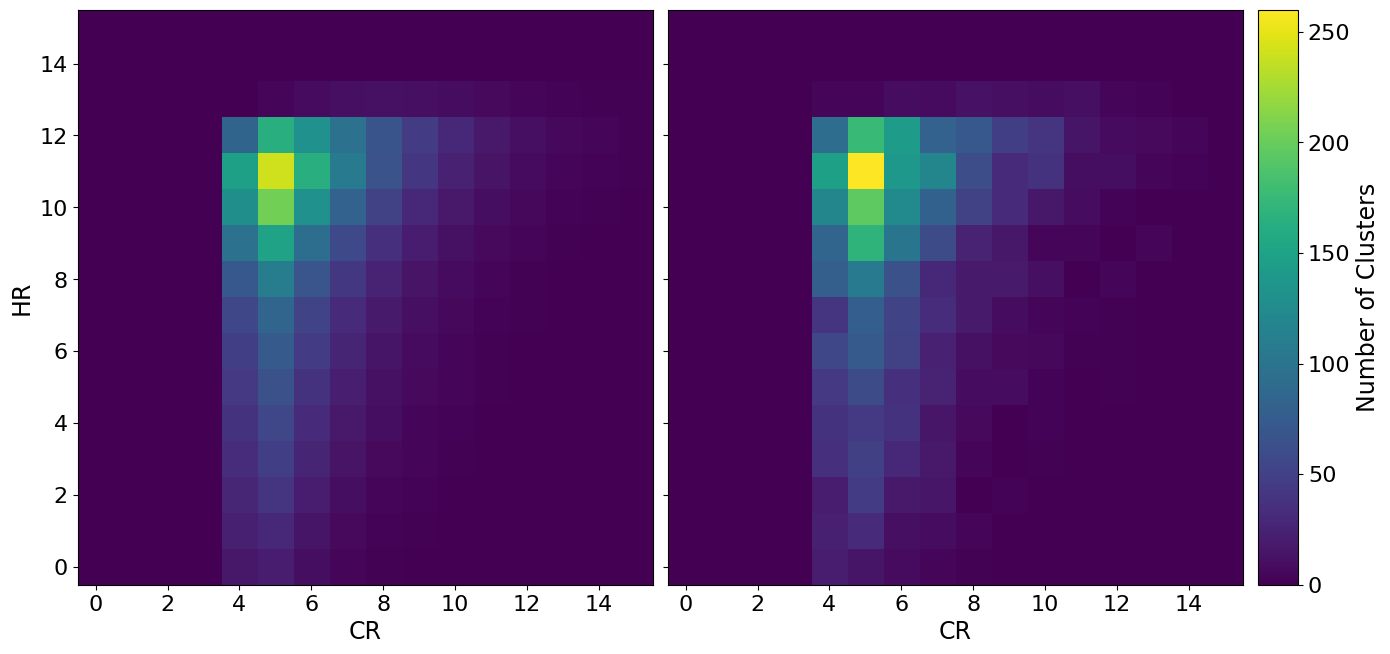}
    \caption{Left: Fiducial XOD for a 1\,000 deg$^2$ survey area (simulated for central values of Table \ref{table:table_parameters}), integrated over the redshift dimension. The XOD has been downsampled to a 16$\times$16$\times$10 resolution, the shape used for the ML model. Right: a particular Poisson realization of the left-hand-side image.}
    \label{fig:XOD_examples_16x16}
\end{center}
\end{figure*}

\begin{figure*}[!th]
\begin{center}

    \includegraphics[height=7.0cm,width=15.2cm,angle=0]{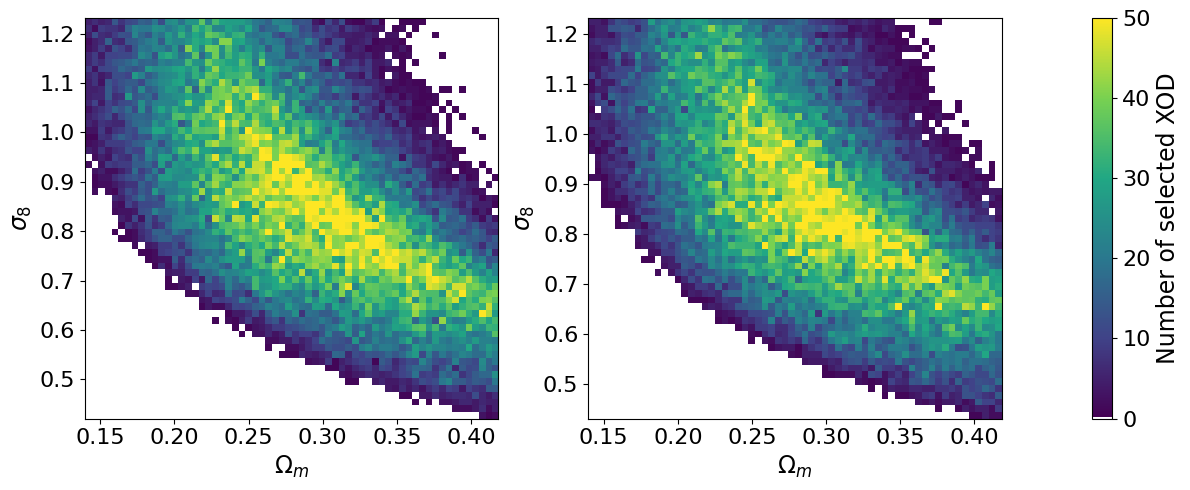}
      \caption{Comparison  of the $\Omega_{\mathrm{m}}$ and $\sigma_8$ distribution for XODs from the D1 and D3  datasets Left: D1 with 8 free parameters. Right: D3  with 6 free parameters, having fixed cluster evolution to be self-similar. Both datasets verify the d$n$/d$z$ test. Each dataset contains 70\,000 XODs. The white zones show the regions excluded by each selection. Fixing cluster evolution does not significantly restrict the a priori range of possible cosmological values. }
      \label{fig:prior_distribution}
\end{center}
\end{figure*}

The CR-HR-$z$ XODs are defined as 64$\times$64$\times$10 arrays for the CR, HR, and $z$ dimensions, respectively (Fig.\,\ref{fig:XOD_examples_64x64}) and are later compressed to a 16$\times$16$\times$10 resolution used for the SNPE ML model (Fig.\,\ref{fig:XOD_examples_16x16}).
To create an XOD, firstly, a set of two cosmological and six scaling relation parameters are drawn from the allowed parameter space (Tab.\,\ref{table:table_parameters}) based on random-uniform prior distributions. 
We then compute the number of halos in the survey volume for a 1\,000 deg$^2$ area, derive the CR for each halo, and apply the CR selection, to include or not the halo in our simulated detected sample. 
Because the analytical derivation of the dn/dM/dz distribution is computer-time demanding, we apply an intermediate test to exclude a priori unphysical combinations of the eight free parameters - this has a high probability of occurring given that the parameters are drawn independently at random. 
For this, we use the fiducial XOD based on the observed cluster distribution in our Universe.
We calculate the number of clusters that pass the CR cut in the second redshift bin (i.e. the most populated one). If this number is incompatible within  3$\sigma$ with the real-life value (i.e. that of our fiducial diagram), the combination is discarded and we consider another set of eight parameters. If this number is compatible with the observed cluster counts, we compute the number of clusters that pass the CR cut in the fourth redshift bin; we perform the same comparison as for the second bin and, similarly, decide whether to keep the set of eight parameters. We perform this test every two redshift bins to validate a parameter combination as plausible quickly. We stop when we have identified  70\,000 combinations of the eight parameters that verify the d$n$/d$z$ constraint for every two bins over the full redshift range.  We find that this only occurs for 1.7\% of the random combinations. The entire process takes about 12 hours to simulate, running in parallel on $\sim$\,110 cores. The sample size of 70\,000 was arbitrarily chosen and later proved to be a sufficiently large number to ensure good performance (Sec.\,\ref{sec:training_data_size}). Fig.\,\ref{fig:prior_distribution} (left) shows the distribution of the $\Omega_{\mathrm{m}}$ and $\sigma_8$ values of the 70\,000 d$n$/d$z$ accepted XODs.

\subsection{Modelling the impact of survey area and of stochastic noise}
\label{sec:Noise_and_survey_area}

In average, the number of detected clusters scales proportionally to the survey area, but the true number of clusters in each XOD pixel is affected by shot noise.
(In all this study, we neglect the effect of sample variance on the cluster number counts). We describe below, how noise is implemented in the diagrams.

The XODs are first simulated for a 1\,000 deg$^2$ survey area with a 64$\times$64$\times$10 resolution, for a given set of cosmological + cluster physics parameters. The d$n$/d$z$ 3$\sigma$ test is applied on this realisation. If the XOD passes the test, the pixel values are multiplied 10, to obtain the XOD for 10 000 deg$^2$.
After that, the XODs are downsampled to the 16$\times$16$\times$10 resolution used for the ML model and we apply a Poisson noise model on every XOD pixel with the Numpy \citep{harris2020array} Python \citep{10.5555/1593511} library. Fig.\,\ref{fig:XOD_examples_16x16} shows an XOD before noising (left) and after the noise (right) at a 16$\times$16$\times$10 resolution. 


In addition, for the purpose of investigating the limits and possible biases of our modelling, we consider two other - unrealistic - survey sizes: 100 000 deg$^2$ and 'infinite'. The infinite realisation is not affected by noise, while the noising of 100 000 deg$^2$ follows the same principles as that of 1 000 deg$^2$ and 10 000 deg$^2$, meaning that the noise in the 100 000 deg$^2$ is equivalent to what noise we would receive if we had such a survey even though it is unrealistic in a real case scenario. We note that since the d$n$/d$z$ test is applied on the 1 000 deg$^2$ realisation only, it is more permissive for the parameter sets compatible with the 10 000 deg$^2$ and 100 000 deg$^2$ surveys.


\subsection{Datasets}
\label{sec:Datasets}

In addition to the main dataset described previously, we also created three other datasets of XODs (Tab.\,\ref{tab:datasets}) to examine how the d$n$/d$z$ selection and how decreasing the number of free parameters by fixing cluster evolution parameters at fiducial values impacts the final accuracy of the cosmological predictions.

\begin{figure*}[!th]
\begin{center}
    \includegraphics[scale=0.39]{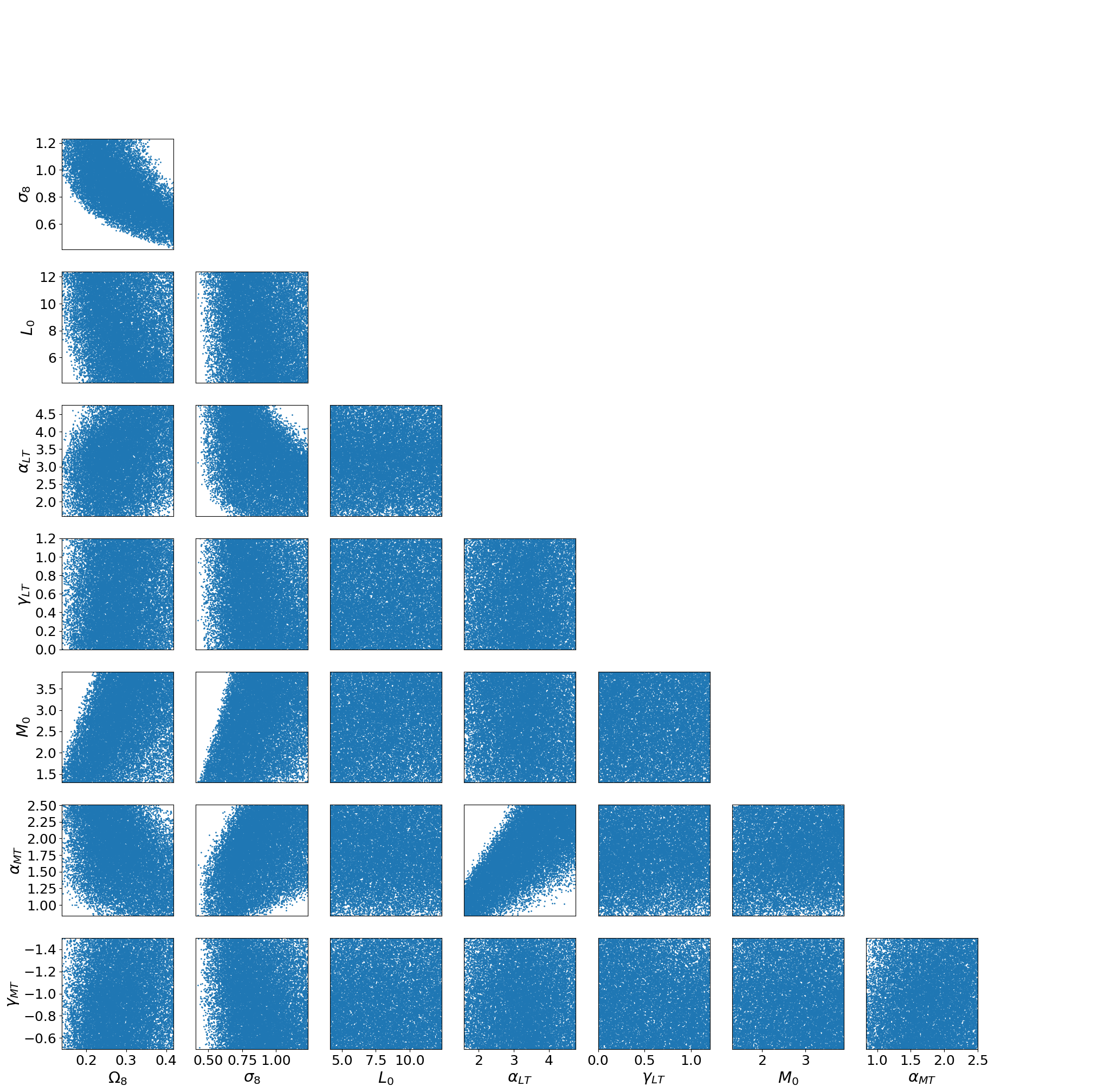}

      \caption{Prior distribution of the two cosmological and six scaling relations parameters after d$n$/d$z$ subselection for D1 dataset. The blue dots represent parameters of XODs that were accepted by the d$n$/d$z$ selection on this dataset. We can see that the d$n$/d$z$ selection flags forbidden regions compared to the initial random uniform selection from this parameter space.
      }
      \label{fig:parameter_distribution}
\end{center}
\end{figure*}

The main dataset for this work is the 70\,000 d$n$/d$z$ accepted XODs created by the steps described previously (D1 dataset). The distribution of its parameters is shown in Fig.\,\ref{fig:parameter_distribution}. The distribution of the parameter values in the dataset shows some forbidden regions in the parameter space, because of the d$n$/d$z$ selection.

The second  XOD dataset inhibits the d$n$/d$z$ subselection (D2 dataset). We do not show the corresponding parameter distribution as it is just the input random uniform one.  

\begin{figure*}[!th]
\begin{center}
    \includegraphics[scale=0.32]{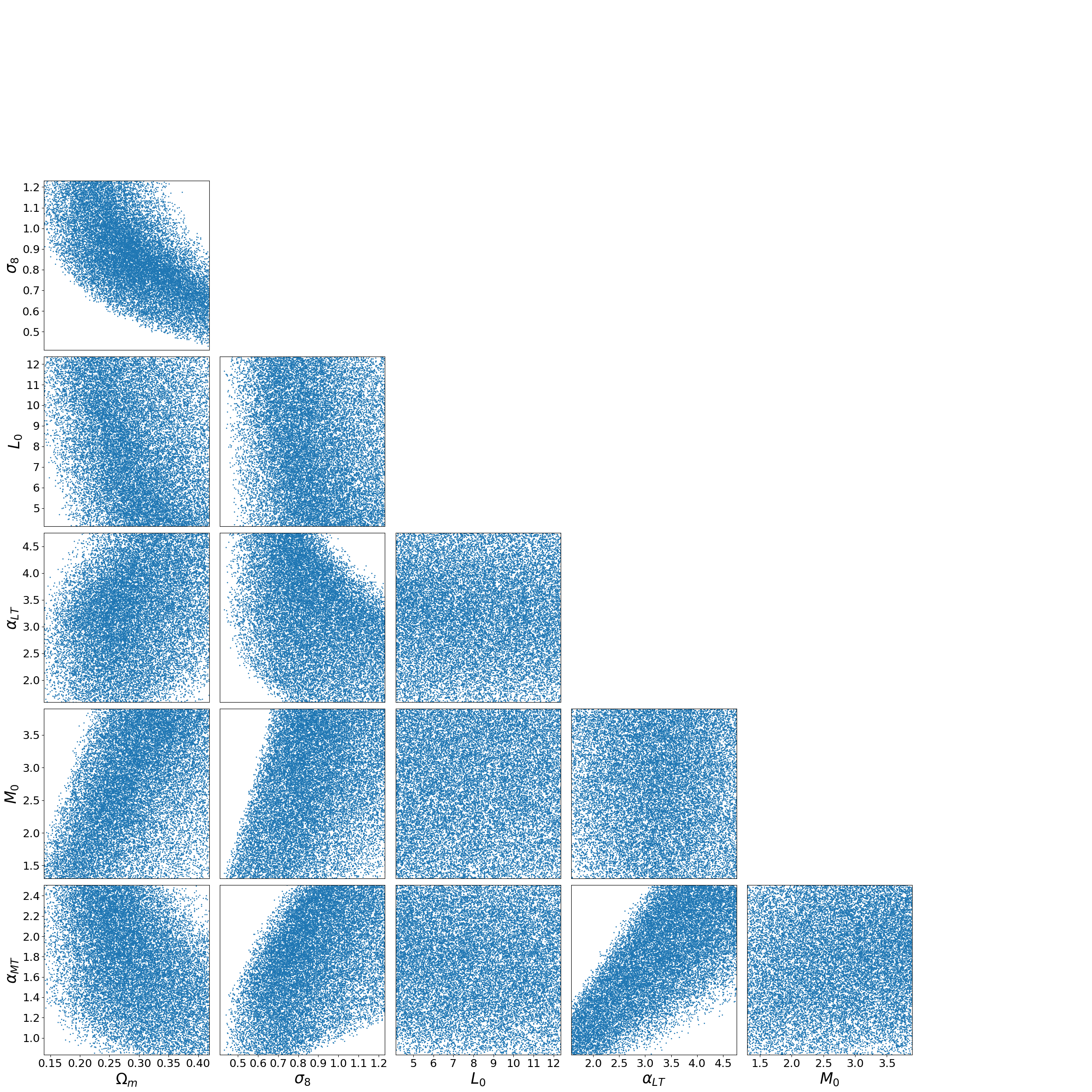}

      \caption{Prior distribution having two cosmological and four scaling relations parameters after d$n$/d$z$ subselection for the ResNet4 training dataset with fixed $\gamma_{MT}$ and $\gamma_{LT}$ scaling relation parameters (D3 dataset). The data points represent XODs accepted in this dataset by the d$n$/d$z$ selection. 
      There are no new forbidden regions compared to the distribution of the D1 dataset (Fig.\,\ref{fig:parameter_distribution}).
      }
      \label{fig:parameter_distribution_fixed_gammas}
\end{center}
\end{figure*}

The third XOD dataset uses the d$n$/d$z$ subselection and, in addition, assumes that cluster evolution is self-similar (i.e. $\gamma_{LT}$ and $\gamma_{MT}$ are fixed at their central fiducial values). this is the D3 dataset (Tab.\,\ref{table:table_parameters}). Fig.\,\ref{fig:parameter_distribution_fixed_gammas} shows the distribution of this dataset's d$n$/d$z$ accepted XOD parameters. 

The last data set (D4) inhibits the d$n$/d$z$ subselection and assumes cluster self-similar evolution.

Thanks to these four data sets, we can investigate the effect of restricting the possible ranges of XOD, by applying well-justified constraints on the cluster scaling relation coefficients, otherwis totally random : (i) the XOD must verify the the observed detected-cluster redshifty distribution and (ii) clusters evolve self-similarly. 

 The datasets are listed in Tab.\,\ref{tab:datasets}. They all contain 70\,000 XODs.

\begin{table}[h]
\caption{The four datasets used for training the networks. Column 2 indicates whether the coefficients of  the cluster scaling relations were selected such as to  match the observed redshift distribution. Column 3 shows whether cluster evolution was fixed to self-similar.  
Each dataset contains 70\,000 XODs. 
}    
\label{tab:datasets}      
\begin{tabular}{ccc}
  Acronym & d$n$/d$z$ & $\gamma_{MT}$ \& $\gamma_{LT}$ \\
  \hline
  D1    & Yes  & Free  \\
  D2    & No   & Free  \\
  D3    & Yes  & Fixed  \\
  D4    & No   & Fixed  \\
    
\hline
\end{tabular}
\end{table}


\section{Sequential Neural Posterior Estimation}
\label{sec:Method}

Simulation-based inference (SBI) is a broad domain encompassing various models. We use the Sequential Neural Posterior Estimation model, SNPE \citep{Papamakarios2016}, from the Python \texttt{sbi} package \citep{tejero-cantero2020sbi}. This model uses neural density estimation (NDE) techniques \citep{Ismail1998} to conduct likelihood-free inference. In this section, we give a non-exhausting basic overview of the SNPE method, and we kindly refer interested readers for more in-depth details to \citep{Papamakarios2016}.

The SNPE and SBI methods, in general, usually work with a compressed version of the data. This is done to preserve a significant amount of relevant information about the data while reducing the dimensionality of the data as much as possible. Compressing the data also keeps the data size for the inference manageable.

We compress our XODs by training a small ResNet model with two resolution levels and two ResNet blocks per level trained under mean square error regression loss as a regressor to estimate the $\Omega_{\mathrm{m}}$ and $\sigma_8$ from the XODs. These estimates are our new parameter space. The XODs enter the SNPE method, which is represented as their compressed version provided by our trained ResNet model. 
For a detailed description of the ResNet architecture, we refer interested readers to \citep{He2015}.

We aim to estimate the posterior probability distribution of the $\Omega_{\mathrm{m}}$ and $\sigma_8$ for a given XOD. We want to get the posterior probability distribution $p(\theta\,|\,x=x_0)$, where $x_0$ corresponds to our target XOD diagram. The posterior probability distribution is proportional as

\begin{equation}
    p(\theta\,|\,x=x_0) \propto p(x = x_0\,|\,\theta) p(\theta)
\end{equation}

\noindent where $p(\theta)$ is our prior and $p(x = x_0\,|\,\theta)$ is the likelihood of our simulator model. In this sense, the simulator model is the pipeline simulating XODs plus the regressor neural network compressing them into the new parameter space. The data, $x$, in our work, represents the compressed version of the XOD.  

With the simulation-based approach of the SNPE, we avoid the explicit computation of the likelihood and its necessary assumptions. The SNPE implements a Mixture Density Network (MDN) \citep{Bishop1994} trained to output a posterior probability distribution as a mixture of 10 Gaussian components. The MDN consists of feed-forward neural networks trained to compute the components' mixing coefficients, means, and covariance matrices. In this work, we use the SNPE-a \citep{Papamakarios2016}. After the training procedure, the MDN models the posterior distribution. 

In our analysis, we used the MDN in a single Gaussian component regime, outputting the posterior probability as a Gaussian to better compare our results with the traditional approach of the Fisher analysis, which predictions are Gaussian.

Our XOD simulations are not sampled from a simple prior. We estimate the empirical proposal prior distribution of the simulations and account for it in the NDE procedure to be able to set our desired prior.

\section{Results}
\label{sec:Results}

\begin{figure*}[!th]
\begin{center}
    \includegraphics[height=7.5cm,width=7.5cm,angle=0]{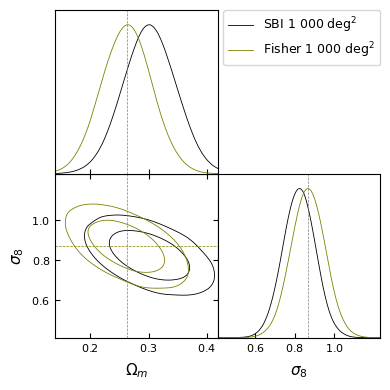}
    \includegraphics[height=7.5cm,width=7.5cm,angle=0]{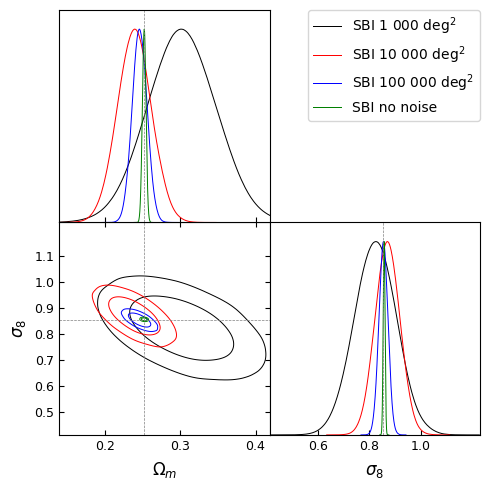}
    \caption{Left: Comparison of the NDE's cosmological prediction with Fisher prediction on the same target XOD corresponding to 1\,000 deg$^2$ survey area. Right: Comparison of the NDE's cosmological results for four different survey areas. The NDE's was trained on our D1 dataset, with d$n$/d$z$ selection and free $\gamma$ parameters. The y-axis in the 1D posterior probability distribution plots indicates the probability density at each parameter value shown on the x-axis.
    Each 1D probability density function is normalized in a way that the total area under its curve equals 1. }
    \label{fig:fisher_prediction}
\end{center}
\end{figure*}

In this section, we present the main results of this work as an estimation of cosmological parameters, $\Omega_{\mathrm{m}}$ and $\sigma_8$ for a target XOD for a survey size of 1\,000 deg$^2$. The dataset used for this experiment is the D1 (see Sec.\,\ref{sec:Datasets}). Fig.\,\ref{fig:fisher_prediction} shows the neural density estimator's cosmological prediction based on this data for a target test XOD. We compare the NDE's prediction with a well-established method, a Fisher analysis, focusing on the same setting as the D1 dataset, with d$n$/d$z$, and including evolution parameters. We compute the Fisher information matrix on the fiducial model parameters. Here, we note an XOD as a set of bins $\lambda_i(\theta)$, with $i$ running over $z$--CR--HR bins, and its likelihood $\mathcal{L}$ can be expressed, assuming that the bins are independent and follow a Poisson distribution, as:

\begin{equation}
    F_{\alpha\beta} = \sum_i \frac{1}{\lambda_i}
                       \frac{\partial \lambda_i}{\partial \theta_\alpha} 
                       \frac{\partial \lambda_i}{\partial \theta_\beta}
.\end{equation}

The derivatives are numerically computed following the methodology presented in \cite{Cerardi2023}, ensuring the output's stability. As the Fisher matrix is computed around a particular point in the parameter space, it does not take account of any proposal distribution. Thus, to represent the effect of the d$n$/d$z$ selection, we compute the covariance matrix of the d$n$/d$z$ training sample and invert it to obtain the Fisher Proposal matrix. Naturally, this induces a loss of information as it only keeps the multi-variate Gaussian signal in the proposal distribution, but it is sufficient for the test we intend to perform here. The Fisher matrices can be added together and then inverted to recover the constraints on $\Omega_{\mathrm{m}}$ and $\sigma_8$. We show the corresponding ellipse in Fig.\ref{fig:fisher_prediction} and an estimated posterior from the NDE for an XOD with its $\Omega_{\mathrm{m}}$ and $\sigma_8$ near the fiducial model. The contours agree well, although the Fisher contours are slightly broader. This could be caused by (i) the prior information drawn from the proposal information being degraded and (ii) the Fisher computation only reflecting the properties of the local derivatives. 
While the Fisher prediction is anchored to the fiducial model and then well-centered, it is normal that the noisy realization of the target diagram offsets the SBI contours. The general agreement between the two constraining methods allows us to increase our confidence in the main results of this paper.

\section{Discussion}
\label{sec:Discussion}

In this section, we first discuss the impact of the size of the training sample on the neural network performances. Next, we show what levels of accuracy can be expected as a function of the survey area. We discuss the effect of the d$n$/d$z$ selection and of the self-similar evolution assumption.  Lastly, we discuss which steps would be taken when using this method on a real observed XOD instead of a simulated one.


\subsection{Training data size}
\label{sec:training_data_size}

In deep learning studies, the larger the training set, the better in principle. 
However, simulating XOD is expensive, we thus investigate the trade-off between the size of the training samples and the returned cosmological accuracy.
Our neural network training consists of two steps (ResNet and NDE) and we thus have to validate the training data size for each of them. 
First,  the ResNet regressor compresses the XODs in a new parameter space, and second, the NDE estimates the posterior probability distribution of $\Omega_{\mathrm{m}}$ and $\sigma_8$ for a given ResNet-compressed XOD. To do this, we fix the training size of one network on a larger volume while varying the size of the other. The results are presented in the next two figures.

\begin{figure*}[!th]
\begin{center}
    
    \includegraphics[height=5.0cm,width=8.5cm,angle=0]{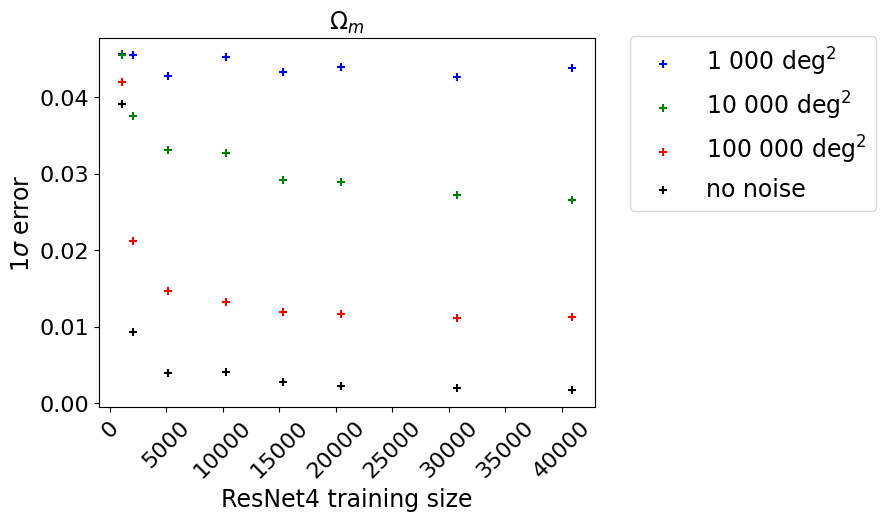}
    \includegraphics[height=5.0cm,width=8.5cm,angle=0]{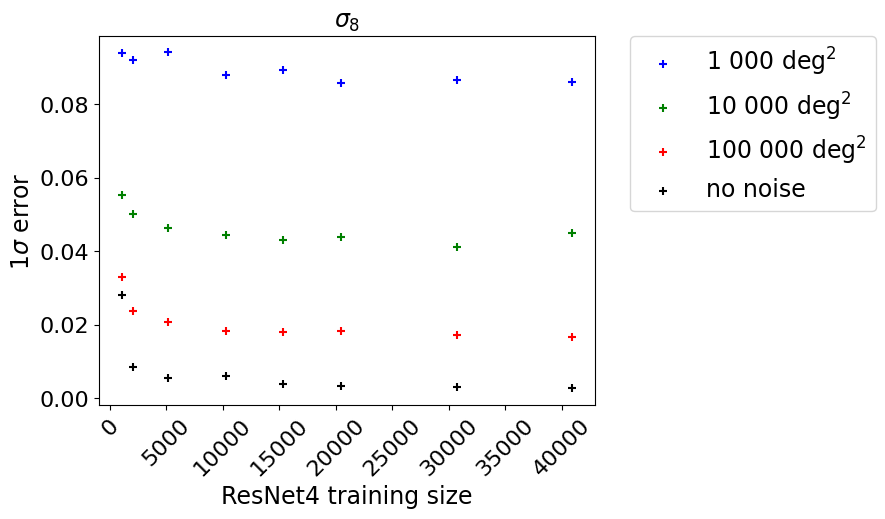}
      \caption{Amplitude of the 1 $\sigma$ errors as a function of the ResNet's training size, for a training size of the NDE set fixed at 10\,000 XODs. We can see that the precision does not significantly improve after 20\,000 training XODs.}
      \label{fig:ResNet4_training_sizes_all_in_one}
\end{center}
\end{figure*}

Fig. \ref{fig:ResNet4_training_sizes_all_in_one} shows how the accuracy on the final prediction varies with the ResNet's training dataset size while keeping the NDE's training size set as 10\,000 XODs. The quality of the final prediction does not significantly improve beyond 10\,000 training samples. 

\begin{figure*}[!th]
\begin{center}
    
    \includegraphics[height=5.0cm,width=8.5cm,angle=0]{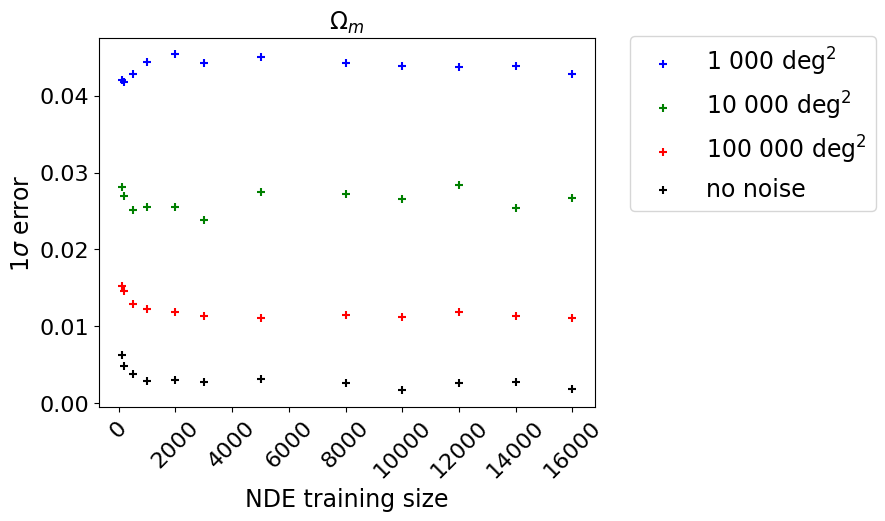}
    \includegraphics[height=5.0cm,width=8.5cm,angle=0]{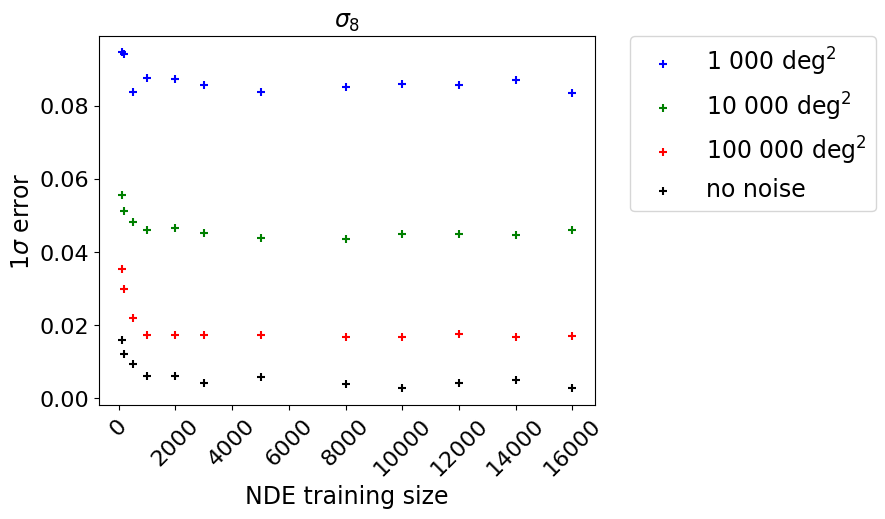}
      \caption{Amplitude of the 1 $\sigma$ errors as a function of the NDE's training size, for a fixed training size of the ResNet set at 40\,960 XODs. We can see that the NDE already performs well with a few hundred XODs as a training size. This was to be expected as it was intentionally developed to work in this way. Even though working already with smaller training XODs, we conservatively decided to use 10\,000 XODs as our final training sample, given that we already simulated 70\,000 XODs.}
      \label{fig:SNPE_training_sizes_all_in_one}
\end{center}
\end{figure*}

Fig. \ref{fig:SNPE_training_sizes_all_in_one} shows how the error on the final prediction varies with the NDE's training dataset size while keeping the ResNet training size set to 40\,960 XODs. 

The quality of the final prediction is more sensitive to the ResNet's training size, while the NDE can already perform well even with a few hundred training examples. This is expected because \citet{Papamakarios2016} designed the NDE to perform well even with a few hundred training samples. Based on these results, we concluded that our initial guess of creating 70\,000 XODs per dataset is enough and that we do not need to enlarge the sample. We set as default values 20\,480 for the ReNet4's training size and 10\,000 XODs for the NDE's training size. If not stated otherwise, the results presented in this paper are always produced with these training sizes.

\subsection{Testing the NDE's calibration}
\label{sec:SNPE_calibration}
\begin{figure*}[!th]
\begin{center}
    
    \includegraphics[height=7.5cm,width=7.5cm,angle=0]
    {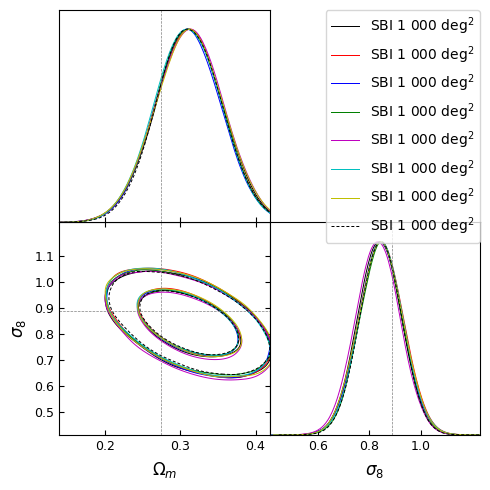}
    \includegraphics[height=7.5cm,width=7.5cm,angle=0]
    {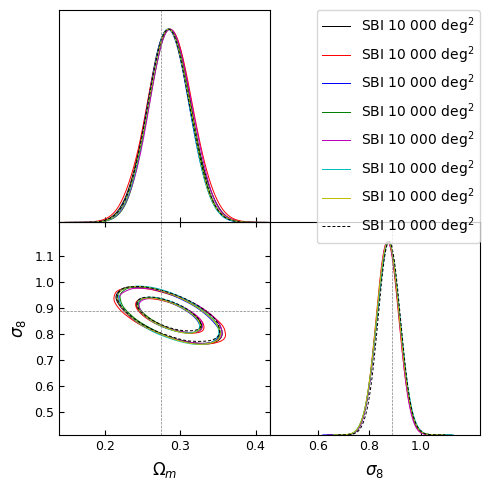}
    \includegraphics[height=7.5cm,width=7.5cm,angle=0]
    {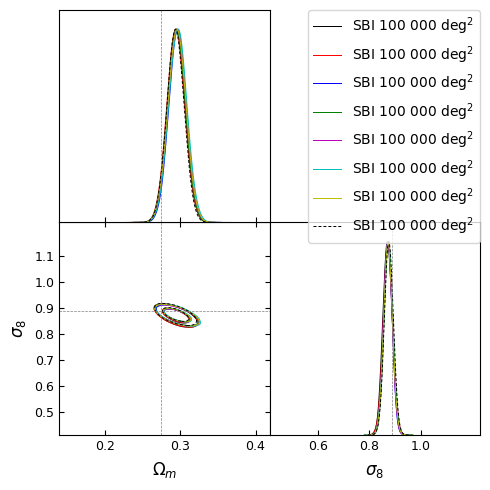}
    \includegraphics[height=7.5cm,width=7.5cm,angle=0]
    {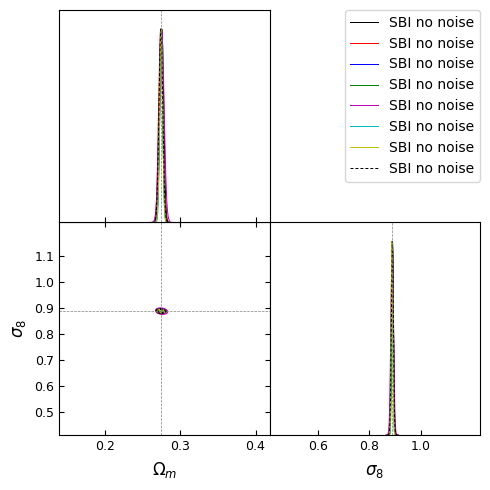}

      \caption{Calibration tests of the density estimator on four different survey areas. Each contour represents an output of a newly re-trained NDE with the same training parameters. The only difference is in the seeds used for generating random numbers, e.g., initialization of weights in the neural network. The tests are done for the same testing XOD. We can see that the NDE is well-trained in that regard it produces consistent results and does not show large deviations only due to small differences in the initialization of its neural networks.}
      \label{fig:results_SNPE_no122_calibration}
\end{center}
\end{figure*}

To test whether the neural density estimator is well-trained and calibrated, we re-train it eight times with the same settings of all hyperparameters, and let it make predictions for the same target XOD. This is done for all four survey areas. In these tests, its training data size is set at 10\,000, and the XODs are always compressed with the same ResNet of the corresponding survey size that is not re-trained. The only difference between these tests is the seed used to draw random numbers from distributions when initializing the NDE's network's layers. If the ML model is adequately trained, it will not have any significant deviations from its predictions when re-trained in this fashion. Fig.\,\ref{fig:results_SNPE_no122_calibration} shows four calibration tests for four different models, each trained for a different noise level represented as a survey size. We can see that the results are very consistent.

\subsection{Survey area }
\label{Survey area - Noise}

As described in Sec\,\ref{sec:Noise_and_survey_area}, we are working with four different survey sizes, i.e., noise levels in the XOD pixels: namely 1\,000\,deg$^{2}$, 10\,000\,deg$^{2}$, 100\,000\,deg$^{2}$, and an infinite survey area (no noising at all). Even though the last two settings are 'thought experiments',  they are potentially useful in understanding possible biases in our methodology. We perform these tests to see what levels of accuracy we can expect for a different survey area and to probe the effect of noise on the accuracy of our cosmological predictions. 

\begin{figure*}[!th]
\begin{center}

    \includegraphics[height=7.5cm,width=7.5cm,angle=0]{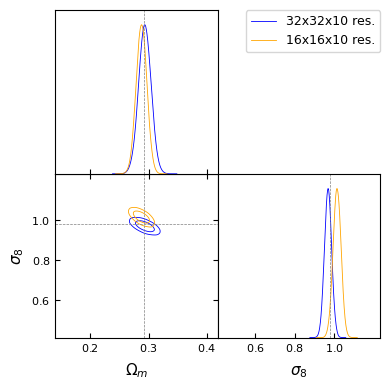}
    \includegraphics[height=7.5cm,width=7.5cm,angle=0]{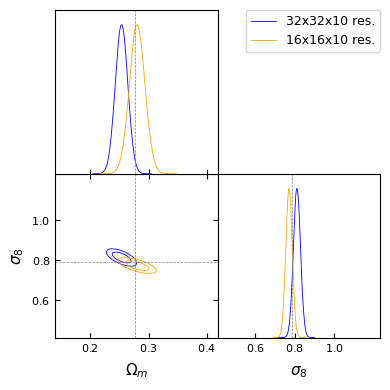}

      \caption{Tests of the accuracy of our cosmological predictions based on the resolution of the XODs in the case of 100\,000 deg$^2$ survey area for two target testing XODs T1 (left) and T2 (right). The XOD resolution is color-coded.}
      \label{fig:resolution_test_100k_deg2}
\end{center}
\end{figure*}

\begin{figure*}[!th]
\begin{center}
    
    \includegraphics[height=7.5cm,width=7.5cm,angle=0]{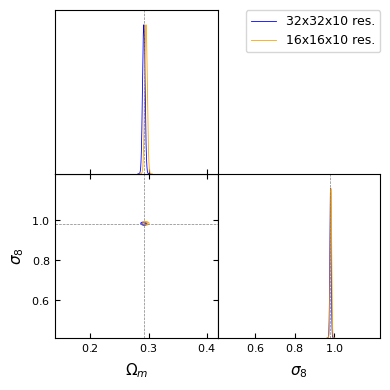}
    \includegraphics[height=7.5cm,width=7.5cm,angle=0]{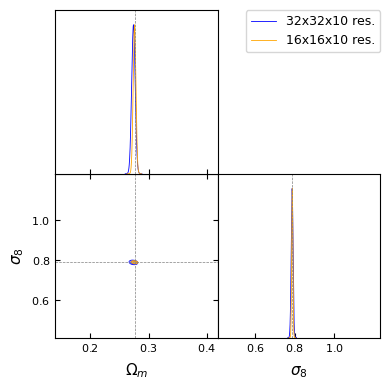}
    
      \caption{Tests of the accuracy of our cosmological predictions based on the resolution of the XODs in the case of no Poisson noise (infinite deg$^2$ survey area) for two target testing XODs T1 (left) and T2 (right). The XOD resolution is color-coded.}
      \label{fig:resolution_test_infinite_deg2}
\end{center}
\end{figure*}

Fig.\,\ref{fig:fisher_prediction} (right) shows how, within our framework, the accuracy of the cosmological parameters depends on the survey size. Tab.\,\ref{table:final_results_errors_noise_levels} summarises the cosmological constraints. Considering the first three columns, it appears that the cosmological accuracy roughly scales as the square root of the area, that is, as the square root of the number of clusters, which is sound. We observe that no noise at all in the XOD (infinite area) still results in non-zero errors. 
We may interpret this as a possible limitation of the numerical accuracy when constructing the XODs. Alternatively, it could also be due to keeping the same 16$\times$16$\times$10 resolution for all noise levels. Improving the resolution in the case of an infinite survey area should lead to smaller errors in this regime. We tested this hypothesis by re-training the ML models for a 32$\times$32$\times$10 XOD resolution. Fig.\,\ref{fig:resolution_test_100k_deg2} shows the final posterior probability cosmological predictions for ML models trained on 32$\times$32$\times$10 XOD resolution (blue) and the standard 16$\times$16$\times$10 XOD resolution (yellow) for two target XODs (left and right figures). Fig.\,\ref{fig:resolution_test_infinite_deg2} shows the same but for a setting with no Poisson noise (infinite deg$^2$). Tab.\,\ref{table:resolution_results_errors} shows the relative 1\,$\sigma$ accuracy of these predictions. We can see that increasing the resolution does not significantly change the accuracy of our cosmological predictions. For the T1 XOD, the relative 1\,$\sigma$ has a bit better accuracy for the 16$\times$16$\times$10 resolution in the case of 100\,000deg$^2$ survey area and only for $\Omega_{\mathrm{m}}$. However, we observe an opposite trend for the T2 target XOD, which had a bit better accuracy on the $\Omega_{\mathrm{m}}$ but for the case of 32$\times$32$\times$10 resolution of 100\,000\,deg$^2$ survey area. For the T2 target XOD, there was also a small improvement in the accuracy of our cosmological prediction for the 16$\times$16$\times$10 resolution in the case of no noise at all (infinite deg$^2$ survey area). Based on these results, we conclude that increasing the resolution to 32$\times$32$\times$10 does not play a significant role in the accuracy of our cosmological predictions, when shot noise becomes negligible. The plausible interpretation is that, given that we work at the CR level, cluster spectral features (in particular emission lines) are somewhat erased by the rather low XMM-EPIC spectral resolution, containing itself sharp discontinuities\footnote{https://xmm-tools.cosmos.esa.int/external/
\\ 
xmm\_user\_support/documentation/uhb/effareaonaxis.html}. This may cause the observed remaining weak degeneracy.

\begin{table}[h]
\caption{This table shows the relative 1\,$\sigma$ errors of our resolution tests for two target XODs T1 and T2 for two different XOD resolutions, 32$\times$32$\times$10 and 16$\times$16$\times$10 (the standard resolution) in a case of two different levels of Poissonian noise, corresponding to 100\,000\,deg$^2$ and an infinite deg$^2$ survey areas.}    
\label{table:resolution_results_errors}      
\begin{tabular}{ccc|cc|cc|cc}
   & \multicolumn{4}{c}{$10^5$\,deg$^2$} & \multicolumn{4}{c}{infinite\,deg$^2$}   \\             
   & \multicolumn{2}{c}{32$\times$32$\times$10} & \multicolumn{2}{c}{16$\times$16$\times$10} & \multicolumn{2}{c}{32$\times$32$\times$10} & \multicolumn{2}{c}{16$\times$16$\times$10}\\
   & $\Omega_{m}$  & $\sigma_8$ &  $\Omega_{m}$ & $\sigma_8$ & $\Omega_{m}$  & $\sigma_8$ &  $\Omega_{m}$ & $\sigma_8$   \\
  \hline
T1 & 3.7 & 1.9 & 3.2 & 2.0 & 0.7 & 0.3 & 0.7 & 0.4  \\
T2 & 4.2 & 2.2 & 4.8 & 2.1 & 1.1 & 0.6 & 0.9 & 0.4  \\

\hline
\end{tabular}
\end{table}

\begin{table}[h]
\caption{Relative 1\,$\sigma$ errors (in \%) from  NDE  for a single target diagram. Separate NDE and regressor are trained for each dataset (D1-D4) and each survey area. The results are computed for the same XOD test target, always of the appropriate survey size, simulated with the $\gamma$ evolution parameters fixed at the fiducial values and with the d$n$/d$z$ selection. This choice allowed us to use it fairly for the ML model trained on each of our four datasets.
}    
\label{table:final_results_errors_noise_levels}      
\begin{tabular}{ccc|cc|cc|cc}
          & \multicolumn{2}{c}{$10^3$\,deg$^2$} & \multicolumn{2}{c}{$10^4$\,deg$^2$} & \multicolumn{2}{c}{$10^5$\,deg$^2$} & \multicolumn{2}{c}{infinite\,deg$^2$}   \\             
                     & $\Omega_{m}$  & $\sigma_8$ &  $\Omega_{m}$ & $\sigma_8$ & $\Omega_{m}$  & $\sigma_8$ &  $\Omega_{m}$ & $\sigma_8$   \\
  \hline
D1 & 15.2 & 10.0 & 9.6 & 5.6 & 4.1 & 2.1 & 0.8 & 0.5  \\
D2 & 17.6 & 13.4 & 11.2 & 7.2 & 7.9 & 5.0 & 4.1 & 2.1  \\
D3 & 12.2 & 5.5 & 5.3 & 2.1 & 2.0 & 0.8 & 0.4 & 0.2  \\
D4 & 15.0 & 10.5 & 7.6 & 4.7 & 4.0 & 2.3 & 2.4 & 0.9  \\

\hline
\end{tabular}
\end{table}


\subsection{d$n$/d$z$ subselection}
\label{sec:results_dndz_subselection}

\begin{figure*}[!th]
\begin{center}
    \includegraphics[height=7.5cm,width=7.5cm,angle=0]{sbi_no140_free_g_yes_dndz_training_target_id_9_fix_g_dndz_y.png}
    \includegraphics[height=7.5cm,width=7.5cm,angle=0]{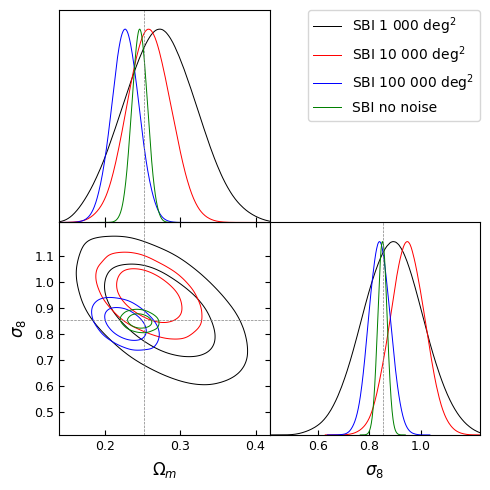}
    
        \caption{Posterior probability distribution, trained on the D1 dataset (d$n$/d$z$ selection, left) and on the D2 dataset (no d$n$/d$z$ selection, right) (Tab.\,\ref{tab:datasets}). These datasets have fixed  $\gamma$ cluster evolution parameters. Four MDNs are trained per dataset each on a different survey area. These results are computed for the same target testing XOD, always of the appropriate survey size, that was produced letting all 8 parameters free and verifying the d$n$/d$z$ selection.}
        \label{fig:results_no_dndz}
\end{center}
\end{figure*}

In this section, we explore how the d$n$/d$z$ selection improves the accuracy of our cosmological predictions. This selection effectively excludes all XODs whose redshift distribution is incompatible with the observations (i.e., here, with the d$n$/d$z$ of our fiducial model) at the 3$\sigma$ level. However, the test is rather permissive, so XODs that pass the d$n$/d$z$ test can show very different combinations of the eight free parameters due to the high level of degeneracy for such a large number of free parameters. Compared to a normal uniform prior distribution, the d$n$/d$z$ restricts the range of possible parameter combinations (Fig.\,\ref{fig:parameter_distribution}), introducing forbidden areas in the parameter space. 

To understand how the d$n$/d$z$ selection improves the final predictions, we compare the ResNet and the NDE trained on the D1 and D2 sets. (see Sec.\,\ref{sec:Datasets}). A separate density estimator must be trained on the corresponding dataset for each survey area. The predictions are made for the same target XOD modulated by the noise level pertaining to the survey area. The results are displayed in Fig.\,\ref{fig:results_no_dndz}. The relative errors are reported in Tab.\,\ref{table:final_results_errors_noise_levels}. We can see that plugging the d$n$/d$z$ selection decreases the size of the error bars on cosmological parameters. This difference is most prominent for smaller noise levels corresponding to larger survey areas. Hence, filtering out unrealistic distributions at the d$n$/d$z$ level is a quick and efficient method to improve the constraining power of this technique.
 

\subsection{Constraining scaling relations - fixing gammas}
\label{sec:fixing_gammas}

\begin{figure*}[!th]
\begin{center}
    \includegraphics[height=7.5cm,width=7.5cm,angle=0]{sbi_no140_free_g_yes_dndz_training_target_id_9_fix_g_dndz_y.png}
    \includegraphics[height=7.5cm,width=7.5cm,angle=0]{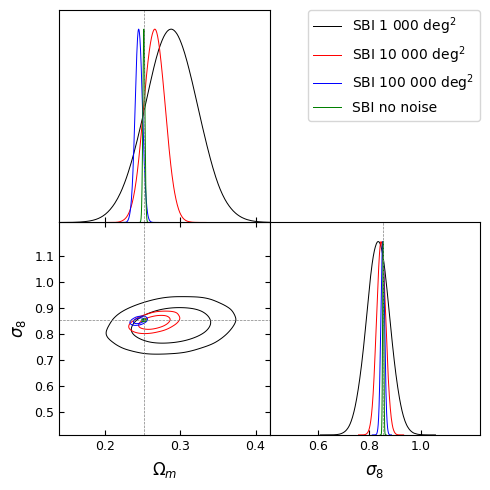}
    
    \caption{Posterior probability distribution estimated trained on D1 dataset (left) and training on D3 dataset (right) (Tab.\,\ref{tab:datasets}). These datasets both have the d$n$/d$z$ selection but differ in the choice of fixing the cluster evolution $\gamma$ parameters on their fiducial values. Four MDNs are trained per dataset each on a different survey area. These results are computed for the same target testing XOD, always of the appropriate survey size, that was produced having all 8 parameters free and with the d$n$/d$z$ selection. We can see that fixing the $\gamma$ parameters improves the accuracy of the NDE's cosmological predictions.}
    
    \label{fig:fixed_gammas}
\end{center}
\end{figure*}

\begin{figure*}[!th]
\begin{center}
    \includegraphics[height=7.5cm,width=7.5cm,angle=0]{sbi_no141_free_g_no_dndz_training_target_id_9_fix_g_dndz_y.png}
    \includegraphics[height=7.5cm,width=7.5cm,angle=0]{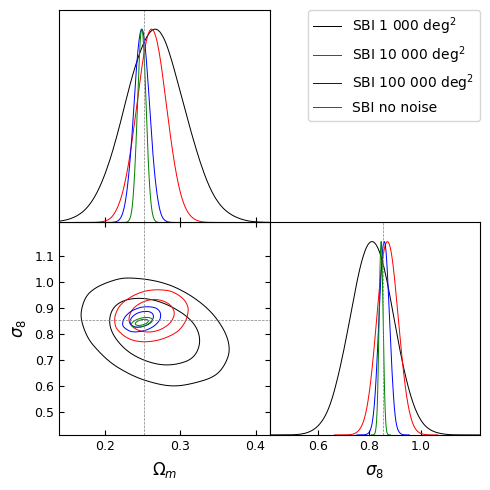}
    
    \caption{Posterior probability distribution estimated trained on D2 dataset (left) and training on D4 dataset (right) (Tab.\,\ref{tab:datasets}). These datasets both come with no d$n$/d$z$ selection but differ in the choice of fixing the $\gamma$ parameters. Four MDNs are trained per dataset each on a different survey area. These results are computed for the same target testing XOD, always of the appropriate survey size, that was produced having all 8 parameters free and with the d$n$/d$z$ selection. We can again observe that using the d$n$/d$z$ selection and fixing the $\gamma$ parameters improves the accuracy of the NDE's cosmological predictions.}
    \label{fig:fixed_gammas_no_dndz}
\end{center}
\end{figure*}

In this section, we investigate how the performance of our cosmological predictions improves when reducing the number of free parameters to six, assuming cluster self-similar evolution  (i.e. $\gamma_{MT}$ and $\gamma_{LT}$ are fixed to their fiducial values). By doing this, the potential degeneracy between cluster physics and cosmology is lowered. 
The test uses data sets D3 and D4.
Fig.\,\ref{fig:fixed_gammas} and Fig.\,\ref{fig:fixed_gammas_no_dndz} show the outcome, compared with the 8-parameter realizations, including or excluding the d$n$/d$z$ selection, respectively. Tab.\,\ref{table:final_results_errors_noise_levels} shows the relative 1\,$\sigma$ errors. 
Clearly, as expected, fixing cluster evolution improves the accuracy of our cosmological predictions.  We also note a change in the direction of the degeneracy.


\subsection{Applying the method to real observations}
\label{sec:real_XOD_use_case}

\begin{figure}[!th]
\begin{center}
    
    \includegraphics[height=7.5cm,width=7.5cm,angle=0]{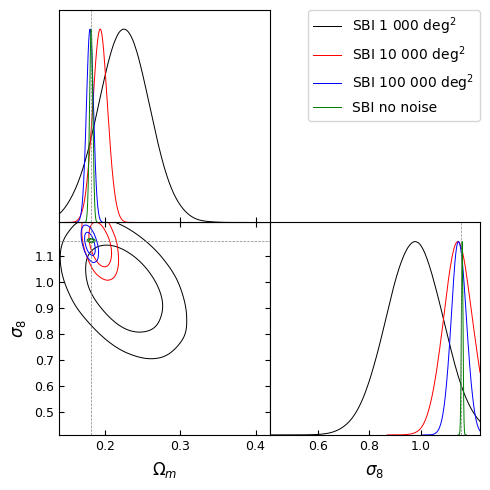}
    
    \caption{This figure shows the NDE's cosmological results for four different survey areas computed for a target XOD with its $\Omega_{\mathrm{m}}$ and $\sigma_8$ values close to the borders of our simulation box. The regressor and density estimator neural networks were trained on our D1 dataset, with d$n$/d$z$ selection and free $\gamma$ parameters.}
    \label{fig:target_at_borders}
\end{center}
\end{figure}

Now, the question is: what steps would be necessary to apply our method to a real observed XOD? So far in the paper, we showcased the method performance on testing XODs which have been inside the parameter range used to create the training simulations (Tab.\,\ref{table:table_parameters}). 
When used on an XOD composed from an observed galaxy cluster sample, with the underlying true $\Omega_{\mathrm{m}}$ and $\sigma_8$ being well within the range of the training simulations parameters, we would see results similar to those demonstrated in this work.
Given the current constraints on $\Omega_{\mathrm{m}}$ and $\sigma_8$ from various probes, we can confidently expect a real XOD to lie well within the boundaries of our study.
However, in principle, we have to address the case where the true cosmological parameters would be located outside of the simulation training range. Since neural regression models are known to extrapolate badly, the posterior estimation would probably fail in this situation. 
This could result in a final estimated posterior being placed towards the edges of the training sample distribution in the direction of the true underlying cosmological parameters.
If we obtained results like this on a real observed XOD, we would need to create a second training sample, now centred on the central values of the estimated posterior probability distribution, and to retrain the neural networks involved, the ResNet regressor as well as the MDN density estimator on this new dataset. These newly trained networks would give us a new estimate of the cosmological parameter posterior probability distribution. We would need to repeat this process until the final estimated posterior would be well-centered within the training sample range. We show such a situation in Fig.\ref{fig:target_at_borders} where the true $\Omega_{\mathrm{m}}$ and $\sigma_8$ cosmological parameters used to simulate this testing XOD are close to the borders of our simulation box.

Other effects will have to be considered when applying this method to a real sample. Firstly, using a CR cut to mimic the selection function is very simplistic and does not correctly render the detection probability within an X-ray survey, which is two-dimensional (flux vs apparent size) \citep{Pacaud2006, Clerc2024}. However, the high flexibility of the ResNet regression models should make the training just as efficient on more complex selection functions. Secondly, measurement errors will have to be included. For X-ray surveys, the number of photons for individual clusters can be as low as a few tens of counts, and hence, the observed CR and HR are noisy estimates of the true quantities. Implementing an error model is not a showstopper for the inference method: basically, this can be modeled by two-dimensional filtering of the XOD \citep{Clerc2012}.

\section{Conclusions}
\label{sec:Conclusions}

This paper is a proof of concept investigating the cosmological constraining power of simulation-based inference utilizing artificial neural networks with a shallow ($\sim$\,200 photons per cluster) X-ray cluster survey using only three observed parameters: the X-ray flux in a given band (XMM-Newton count rate, CR), an X-ray color (XMM-Newton hardness ratio, HR), and the redshift. Our simulated cluster populations are thus represented in this CR-HR-$z$ parameter space, an X-ray observational diagram (XOD).   

The neural networks in our cosmological inference are trained on a large number of simulated XODs. 
To simulate a sample of XODs, we forward model cluster X-ray observables from their mass through scaling relation parameters.  
The scaling-relation parameters are selected randomly from the prior distribution, and they never directly enter our regressor neural network that compresses the XODs nor the density estimator neural network that followingly estimates the posterior probability distribution of the $\Omega_{\mathrm{m}}$ and $\sigma_8$ for a target XOD.



This process allows us to perform a likelihood-free cosmological inference. The size of the posterior is governed by the regressor compression
(limited by the Poison noise in the XODs and the parametrization used to model the CR-HR-z observational quantities).

The performances of our approach are sensible: cosmological constraints scale roughly as the square root of the surveyed area, imposing that the simulated XODs verify the observed d$n$/d$z$ cluster distribution, and fixing the cluster evolutionary parameters leads to better constraints.

The absolute constraining power of our method can be compared to the recent eROSITA cosmological results \citep{Ghirardini2024}. Their cosmological sample has 
5\,259 massive clusters over 12\,791\,deg$^2$; our 1\,000\,deg$^2$ realization yields about 4\,000 clusters. Our predicted uncertainties on $\Omega_{\mathrm{m}}$ and $\sigma_8$ are $\sim$\,2.2 and $\sim$\,4.3 times larger, respectively. This is not surprising because (i) the eROSITA clusters are more massive compared to our 1\,000\,deg$^2$ because our sample has a flux cut based on the XXL analysis, thus having a similar cluster mass distribution, also having $\sim$ 4\,000 clusters on a 1\,000\,deg$^2$ compared to the similar 5\,259 eROSITA clusters over 12\,791\,deg$^2$ and (ii) several 
assumptions are made on the eROSITA cluster scaling relations (from optical data sets) to calibrate the individual cluster masses \citep{Ghirardini2024}.
 
Applying our method to real data would require implementing a two-dimensional selection function in the CR vs cluster-apparent-size plane. Practically, this means that we need to introduce a supplementary scaling relation ($M_{500}$-$R_{\mathrm c}$) in the formalism that produces the XODs; this is not a problem as long as the coefficients of this new relation would be drawn at random, the same way we do with the scaling relations in this study. Modeling the measurement errors would also be needed.

The present study treats the scaling relation coefficients as nuisance parameters; this is a fair approach, but because these parameters amount to six, and we exclude any a priori knowledge on them, there remains a significant degeneracy with the cosmological parameters. This explains why our predicted errors on the cosmological parameters appear rather large.
A much more efficient approach is to create the XOD training sample from numerical simulations spanning a large range of cosmologies and for various feedback assumptions \citep[e.g., the CAMELS simulations,][]{Villaescusa-Navarro2021}. In this case, the number of parameters ruling the cluster's physical properties is drastically reduced (two AGN feedback parameters) 
and, moreover, these parameters should not be considered as nuisance parameters. In the end, the SBI, relying on such a training sample, faces much less degeneracy and delivers the cosmological parameters plus the feedback parameters, which are the truly relevant physics parameters (paper VII, Cerardi et al. in prep).

\begin{acknowledgements}

We are very grateful for Tom Charnock's huge help and contribution to this project in its early stages. 
M. K. and N. W. are supported by the GACR grant 21-13491X. M. K.'s stays in France and Italy were supported by the French Alternative Energies and Atomic Energy Commission internship, the ERASMUS program, and the Italian Government Scholarship issued by the Italian MAECI.

\end{acknowledgements}

\bibliographystyle{aa} 
\bibliography{bibliography}

\begin{appendix} 


\end{appendix} 

\end{document}